\documentclass[letterpaper,onecolumn,10pt,final,cleanfoot,cleanhead]{asme2e}

\usepackage{graphicx} 
\graphicspath{ {Images/} }
\usepackage{epsfig}   
\usepackage{mathptmx} 
\usepackage{times}    
\usepackage{amsmath}  
\usepackage{amssymb}  
\usepackage{units}    
\usepackage{booktabs} 
\usepackage{cite}     
\usepackage{newtxtext, newtxmath}   
\PassOptionsToPackage{normalem}{ulem}
\usepackage{ulem}
\usepackage[hyphens]{url} 
\usepackage{color}
\usepackage{hyperref} 
\usepackage{subfig}   
\usepackage{float}    

\newcommand{\R}{\mathbb{R}}
\newtheorem{proposition}{Proposition}
\newtheorem{assum}{Assumption}
\newtheorem{remark}{Remark}

\confshortname{the ASME 2018 Dynamic Systems and Control Conference}
\conffullname{DSCC2018}
\confdate{September 30-October 3}
\confyear{2018}
\confcity{Atlanta, Georgia}
\confcountry{USA}

\papernum{DSCC2018-9066}
\title{REACHABILITY ANALYSIS FOR ROBUSTNESS EVALUATION OF THE SIT-TO-STAND MOVEMENT FOR POWERED LOWER LIMB ORTHOSES}
\author{Octavio Narvaez-Aroche\thanks{Address all correspondence to this author.}\\
	{\tensfb Andrew Packard}
	\affiliation{Berkeley Center for Control and Identification\\
		Department of Mechanical Engineering\\
		University of California\\
		Berkeley, California, 94720\\
		Email: ocnaar@berkeley.edu, apackard@berkeley.edu
	}	
}

\author{Pierre-Jean Meyer\\ 
	{\tensfb Murat Arcak}
	\affiliation{Department of Electrical Engineering\\
		and Computer Sciences\\
		University of California\\
		Berkeley, California, 94720\\
		Email: pjmeyer@berkeley.edu, arcak@berkeley.edu
	}	
}

\begin{document}
	
	\maketitle
	
	\begin{abstract}
		
		\it{A sensitivity-based approach for computing over-approximations of reachable sets, in the presence of constant parameter uncertainties and a single initial state, is used to analyze a three-link planar robot modeling a Powered Lower Limb Orthosis and its user. Given the nature of the mappings relating the state and parameters of the system with the inputs, and outputs describing the trajectories of its Center of Mass, reachable sets for their respective spaces can be obtained relying on the sensitivities of the nonlinear closed-loop dynamics in the state space.
			These over-approximations are used to evaluate the worst-case performances of a finite time horizon linear-quadratic regulator (LQR) for controlling the ascending phase of the Sit-To-Stand movement.}
		
	\end{abstract}
	\section{INTRODUCTION}
	
	Powered Lower Limb Orthoses (PLLOs) are medical devices worn in parallel of the legs that must work in synchrony with their users to assist standing and/or walking. State of the art PLLOs for people with paraplegia ($\approx$114,000 individuals in the USA \cite{NSCISC2018}) can be safely used for gait training\cite{Baunsgaard2018}, but they have yet to provide full autonomy to perform the Sit-to-Stand (STS) movement, which is the sequence of actions executed for rising from a chair. The STS movement consists of three phases: preparation, ascending and stabilization \cite{Galli2008}. Since a PLLO must ensure safety, regardless of variability of its dimensions from manufacturing, and the weight fluctuations of its user, we aim to analyze the robustness of a controlled PLLO against parameter uncertainty.
	
	In this paper, the robustness is evaluated through the use of reachability analysis, which deals with the problem of computing the set of all possible successors of a system, given its initial state and a set of admissible parameters.
	Since a reachable set can rarely be computed exactly except in simple cases~\cite{asarin1995reachability}, we instead rely on the computation of over-approximations, for which various methods and representations exist, such as ellipsoids~\cite{kurzhanskiy2007ellipsoidal}, polytopes~\cite{chutinan2003computational} or level-sets~\cite{mitchell2000level}.
	The considered approach is based on the results presented in~\cite{meyer2018sampled}, where the computation of interval over-approximations for an uncertain system relies on its sensitivity matrices, i.e.\ the partial derivatives of its trajectories with respect to the uncertain parameters.
	While being inspired by the results in~\cite{xue2017cdc} for the case of systems whose sensitivity matrix is sign-stable over the set of parameters, the strength of the results from~\cite{meyer2018sampled} used in this paper is that it is applicable to any dynamical system whose sensitivity matrix is bounded.
	
	The main objective of our study is to apply this reachability analysis approach to the PLLO, in order to evaluate the worst-case performances of the closed-loop behavior obtained from the finite horizon linear-quadratic regulator (LQR) designed in~\cite{Narvaez-Aroche2018}. Since a proper evaluation of these performances should not be limited to the states, but also include the position and velocity of the Center of Mass (CoM), and the inputs; we extend the method in~\cite{meyer2018sampled} to be able to apply the reachability analysis to static systems such as those defined by an output map of the system or the feedback controller.
	
	We start by reviewing the dynamics of the three-link planar robot used to model the PLLO and its user, the motion planning strategy for obtaining adequate reference trajectories for the ascending phase of a STS movement, and the equations required to solve for the design of the LQR controller. 
	The reachability analysis from~\cite{meyer2018sampled} is then presented for a generic dynamical system along its extension to deal with static systems.
	Finally, these results are applied to the closed-loop PLLO in simulation to assess the robustness of the LQR controller.
	
	Since coordinate aligned boxes play an important role in this study, for $a,b\in\mathbb{R}^n$ we use the notations $a< b$ to mean $a_i< b_i$ $\forall i$ (with similar elementwise definitions for $\leq$, $\geq$, and $>$) and we define an interval of $\R^n$ as $\left[a,b \right]:=\{\xi \in\mathbb{R}^n | a \leq \xi \leq b\}\subseteq\mathbb{R}^n$.   
	\section{DYNAMICS FOR MODELING THE POWERED LOWER LIMB ORTHOSIS AND ITS USER}
	
\begin{figure}
		\centering
		\includegraphics[width=5.5cm]{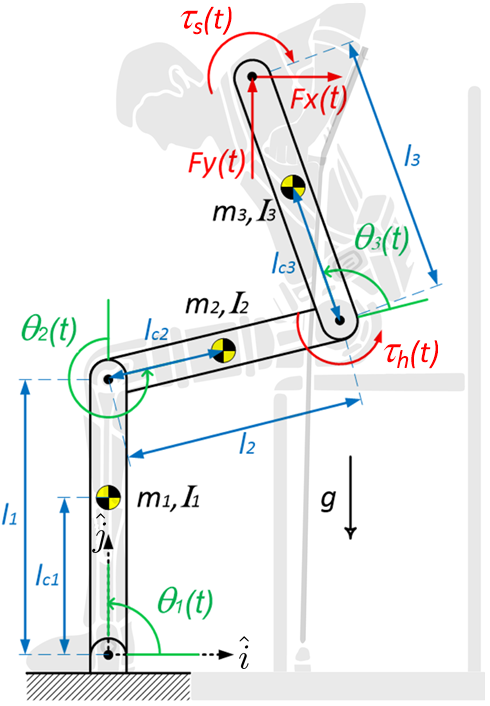}
		\caption{Three-link planar robot for modeling a Powered Lower Limb Orthosis (PLLO) and its user during a Sit-To-Stand (STS) movement.\label{fig:Robot}}
	\end{figure}

	Assuming sagittal symmetry, no movement of the head relative to the torso, and that feet are fixed to the ground, we model the user, crutches and PLLO as a three-link planar robot with revolute joints coaxial to the ankles, knees and hips, as shown in Figure \ref{fig:Robot}. $\theta_{1}$ is the angular position of link 1 (shanks) measured from the horizontal, $\theta_{2}$ is the angular position of link 2 (thighs) relative to link 1, and $\theta_{3}$ is the angular position of link 3 (torso) relative to link 2. The system parameters are the masses of the links $m_1$, $m_2$, and $m_3$; the moments of inertia about their respective CoMs $I_{1}$, $I_{2}$, and $I_{3}$; their lengths $l_1$, $l_2$, and $l_3$; and the distances of their CoMs from the joints $l_{c1}$, $l_{c2}$, and $l_{c3}$. The actuators of the orthosis exert torque $\tau_{h}$ about the hips; while torque $\tau_{s}$, horizontal force $F_{x}$ and vertical force $F_{y}$ capture the inertial and gravitational forces of the arms and loads applied on the shoulders of the user by its interaction with the ground through crutches. There is no actuation at the knees in compliance to the architecture used in the most affordable device in the market for users with complete paraplegia \cite{ExoskeletonReportLLC2018}. 
	
	For notational convenience, denote $ c_{i} :=\cos\theta_{i}\left(t\right) $, $ c_{ij} :=\cos\left(\theta_{i}\left(t\right)+\theta_{j}\left(t\right)\right) $, $ c_{ijk} :=\cos\left(\theta_{i}\left(t\right)+\theta_{j}\left(t\right)+\theta_{k}\left(t\right)\right) $,
	and similarly for $ \sin\left(\cdot\right) $. In terms of the joint angles {\small $\theta=\left[\theta_{1};\;\; \theta_{2};\;\;\theta_{3}\right]$}, input {\small $u=\left[\tau_{h};\;\; \tau_{s};\;\; F_{x};\;\; F_{y}\right]$}, parameters
	{\small{
			\begin{align*}
			p&=\left[m_{1};\;\;m_{2};\;\;m_{3};\;\;I_{1};\;\;I_{2};\;\;I_{3};\;\;l_{1};\;\;l_{2};\;\;l_{3};\;\;l_{c1};\;\;l_{c2};\;\;l_{c3}\right],
			\end{align*}}}
	and{\small \[
		\begin{array}{lll}
		k_{0}\left(p\right):=\left(m_{1}+m_{2}+m_{3}\right)^{-1}, &  & k_{1}\left(p\right):=l_{c1}m_{1}+l_{1}m_{2}+l_{1}m_{3},\\
		k_{2}\left(p\right):=l_{c2}m_{2}+l_{2}m_{3}, &  & k_{3}\left(p\right):=l_{c3}m_{3},
		\end{array}
		\]}the Euler-Lagrange equations of the three-link planar robot in Figure \ref{fig:Robot} can be written, with the aid of the symbolic multibody dynamics package PyDy \cite{Gede2013}, as
	{\small \begin{equation}
		M\left(\theta\left(t\right),p\right)\ddot{\theta}\left(t\right)+F\left(\theta\left(t\right),\dot{\theta}\left(t\right),p\right)=A_{\tau}\left(\theta\left(t\right),p\right)u\left(t\right). \label{eq:EulerLagrange}
		\end{equation}}$M\left(\theta,p\right)\in\mathbb{R}^{3\times3}$, $M\left(\theta,p\right)\succ0$ is the symmetric mass matrix of the system with entries
	{\small{\begin{align*}
			M_{11} & =  I_{1}+I_{2}+I_{3}+l_{c1}^{2}m_{1}+m_{2}\left(l_{1}^{2}+2l_{1}l_{c2}c_{2}+l_{c2}^{2}\right) + m_{3}\left(l_{1}^{2}+2l_{1}l_{2}c_{2}+2l_{1}l_{c3}c_{23}+l_{2}^{2}+2l_{2}l_{c3}c_{3}+l_{c3}^{2}\right)\\
			M_{12} & = I_{2}+I_{3}+l_{c2}m_{2}\left(l_{1}c_{2}+l_{c2}\right) + m_{3}\left(l_{1}l_{2}c_{2}+l_{1}l_{c3}c_{23}+l_{2}^{2}+2l_{2}l_{c3}c_{3}+l_{c3}^{2}\right)\\
			M_{13} & = I_{3}+l_{c3}m_{3}\left(l_{1}c_{23}+l_{2}c_{3}+l_{c3}\right)\\
			M_{22} & = I_{2}+I_{3}+l_{c2}^{2}m_{2}+m_{3}\left(l_{2}^{2}+2l_{2}l_{c3}c_{3}+l_{c3}^{2}\right)\\
			M_{23} & = I_{3}+l_{c3}m_{3}\left(l_{2}c_{3}+l_{c3}\right)\\
			M_{33} & = I_{3}+l_{c3}^{2}m_{3}.
			\end{align*}}}$F\left(\theta,\dot{\theta},p\right)\in\mathbb{R}^{3}$ is the vector of energy contributions due to the acceleration of gravity $g=9.81\left[\nicefrac{m}{s^{2}}\right]$ and Coriolis forces
	{\footnotesize{\[
			F\left(\theta,\dot{\theta},p\right)=\Omega\left(\theta,p\right)\left[\begin{array}{c}
			\dot{\theta}_{1}^{2}\\
			\left(\dot{\theta}_{1}+\dot{\theta}_{2}\right)^{2}\\
			\left(\dot{\theta}_{1}+\dot{\theta}_{2}+\dot{\theta}_{3}\right)^{2}
			\end{array}\right]+g\left[\begin{array}{c}
			k_{1}{\scriptstyle \left(p\right)}c_{1}+k_{2}{\scriptstyle \left(p\right)}c_{12}+k_{3}{\scriptstyle \left(p\right)}c_{123}\\
			k_{2}{\scriptstyle \left(p\right)}c_{12}+k_{3}{\scriptstyle \left(p\right)}c_{123}\\
			k_{3}{\scriptstyle \left(p\right)}c_{123}
			\end{array}\right],
			\]}} with 
	{\footnotesize \[
		\Omega\left(\theta,p\right)=\left[\begin{array}{ccc}
		l_{1}\left(k_{2}{\scriptstyle \left(p\right)}s_{2}+k_{3}{\scriptstyle \left(p\right)}s_{23}\right) & -k_{2}{\scriptstyle \left(p\right)}l_{1}s_{2}+k_{3}{\scriptstyle \left(p\right)}l_{2}s_{3} & -k_{3}{\scriptstyle \left(p\right)}\left(l_{1}s_{23}+l_{2}s_{3}\right)\\
		l_{1}\left(k_{2}{\scriptstyle \left(p\right)}s_{2}+k_{3}{\scriptstyle \left(p\right)}s_{23}\right) & k_{3}{\scriptstyle \left(p\right)}l_{2}s_{3} & -k_{3}{\scriptstyle \left(p\right)}l_{2}s_{3}\\
		l_{1}k_{3}{\scriptstyle \left(p\right)}s_{23} & k_{3}{\scriptstyle \left(p\right)}l_{2}s_{3} & 0
		\end{array}\right].
		\]}
	$A_{\tau}\left(\theta,p\right)\in\mathbb{R}^{3\times4}$ is the generalized force matrix 
	{\footnotesize \[
		A_{\tau}\left(\theta,p\right)=\left[\begin{array}{cccc}
		0 & -1 & -l_{1}s_{1}-l_{2}s_{12}-l_{3}s_{123} & l_{1}c_{1}+l_{2}c_{12}+l_{3}c_{123}\\
		0 & -1 & -l_{2}s_{12}-l_{3}s_{123} & l_{2}c_{12}+l_{3}c_{123}\\
		1 & -1 & -l_{3}s_{123} & l_{3}c_{123}
		\end{array}\right].
		\]}
	\section{MOTION PLANNING}
	
	Biomechanical studies measure the kinematics of the CoM of the human body instead of joint angles to classify and assess dynamic balance of the STS movement \cite{Fujimoto2012}. Therefore, considering $\theta_{2}$, and the position coordinates of the CoM of the three-link planar robot in its inertial frame $\left(x_{CoM},y_{CoM}\right)$, we define $ z:=\left[\theta_{2}; x_{CoM}; y_{CoM}\right]$ and plan the STS motion over the finite time horizon $t\in\left[t_{0},t_{f}\right]$ with reference trajectories {\small \begin{align}\begin{split}
		\hat{\theta}_{2}\left(t\right) & =  \hat{\theta}_{2}\left(t_{0}\right)+\left(\hat{\theta}_{2}\left(t_{f}\right)-\hat{\theta}_{2}\left(t_{0}\right)\right)\Theta_{1}\left(t,t_{f}\right),\\
		\hat{x}_{CoM}\left(t\right)    & =  \hat{x}_{CoM}\left(t_{0}\right)+\left(\hat{x}_{CoM}\left(t_{f}\right)-\hat{x}_{CoM}\left(t_{0}\right)\right)\Theta_{2}\left(t,t_{f}\right),\\
		\hat{y}_{CoM}\left(t\right)    & =  \hat{y}_{CoM}\left(t_{0}\right)+\left(\hat{y}_{CoM}\left(t_{f}\right)-\hat{y}_{CoM}\left(t_{0}\right)\right)\Theta_{3}\left(t,t_{f}\right),
		\end{split}\label{eq:MotionPlanning}\end{align}}where ${\small \Theta_{i}\left(t,t_{f}\right)}$ are polynomial functions satisfying ${\small \Theta_{i}\left(t_{0},t_{f}\right)=0}$ and ${\small \Theta_{i}\left(t_{f},t_{f}\right)=1}$. This rest-to-rest maneuver formulation is taken from \cite{Sira-Ramirez2004}.
	
	Relying on kinematic equations, we showed in \cite{Narvaez-Aroche2017} that for feasible and realistic STS movements excluding the vertical position ($\theta_{1}=\nicefrac{\pi}{2}$, $\theta_{2}=\theta_{3}=0$), a transformation of the form {\small \begin{equation}
		\left[
		\hat{\theta}\left(t\right); \;\; \dot{\hat{\theta}}\left(t\right); \;\; \ddot{\hat{\theta}}\left(t\right)\right]=h\left(\hat{z}\left(t\right),\dot{\hat{z}}\left(t\right),\ddot{\hat{z}}\left(t\right),\hat{p}\right)
		\label{eq:z2theta}\end{equation}}exists; so that once $\dot{\hat{z}}$ and $\ddot{\hat{z}}$ are computed from (\ref{eq:MotionPlanning}), the reference trajectories for the ascending phase in the $z$ space can be mapped into $\theta$ with the nominal values of the parameters $\hat{p}$.
	
	We take a computed torque approach \cite{Slotine1991} for obtaining the reference trajectories $\hat{u}\left(t\right)$. Since the system of equations in (\ref{eq:EulerLagrange}) is underdetermined,  we solve, at every $t\in\left[t_{0},t_{f}\right]$, a control allocation problem \cite{Johansen2013} with the constrained least-squares program {\small \begin{align}
		\hat{u}\left(t\right)= & \:\underset{\xi\in\mathbb{R}^4}{\arg\min} \quad \frac{1}{2}\left\Vert W_{u}\:\xi\right\Vert _{2}^{2}\label{eq:Allocation}\\
		& \mathrm{subject\: to\:}   A_{\tau}\left(\hat{\theta}\left(t\right),\hat{p}\right)\xi=M\left(\hat{\theta}\left(t\right),\hat{p}\right)\ddot{\hat{\theta}}\left(t\right)+F\left(\hat{\theta}\left(t\right),\dot{\hat{\theta}}\left(t\right),\hat{p}\right)\nonumber\\
		& \quad \quad \quad \quad \underline{u}\leq\xi\leq \overline{u},\nonumber
		\end{align}}where $W_{u}\in\mathbb{R}^{4\times4}$ and $\underline{u},\overline{u}\in\mathbb{R}^{4}$ are user-specified weights and box constraints, respectively.
	\section{FINITE TIME HORIZON LQR CONTROLLER}
	
	The Euler-Lagrange equations must be linearized in order to design an LQR controller. Define $x\in\mathbb{R}^6$ as {\small $ x:=\left[\,\theta; \:\dot{\theta}\right] $}, from (\ref{eq:EulerLagrange}), the dynamics of the three-link planar robot are 
	{\small \begin{align*}
		\dot{x}\left(t\right) & = \left[\begin{array}{c}
		\dot{\theta}\left(t\right)\\
		M^{-1}\left(\theta\left(t\right),p\right)\left(A_{\tau}\left(\theta\left(t\right),p\right)u\left(t\right)-F\left(\theta\left(t\right),\dot{\theta}\left(t\right),p\right)\right)
		\end{array}\right]\\
		& =: f\left(x\left(t\right),p,u\left(t\right)\right)
		\end{align*}}With reference state trajectories {\small $\hat{x}\left(t\right):=\left[\hat{\theta}\left(t\right),\dot{\hat{\theta}}\left(t\right)\right]$} from \eqref{eq:MotionPlanning} and\eqref{eq:z2theta}, the state deviation variables $\delta_x\left(t\right)=x\left(t\right)-\hat{x}\left(t\right)$ satisfy {\small \[
		\dot{\delta}_{x}\left(t\right):=f\left(x\left(t\right),p,u\left(t\right)\right)-f\left(\hat{x}\left(t\right),\hat{p},\hat{u}\left(t\right)\right),
		\]}which can be approximated with a first order Taylor series expansion of $f\left(x\left(t\right),p,u\left(t\right)\right)$ about $\hat{x}\left(t\right)$, $\hat{p}$ and $\hat{u}\left(t\right)$: {\small \begin{align}
		\dot{\delta}_{x}\left(t\right) & \approx \left.\frac{\partial f\left(x,p,u\right)}{\partial x}\right|_{\scriptsize {\begin{array}{l}
				x=\hat{x}\left(t\right)\nonumber\\
				p=\hat{p}\\
				u=\hat{u}\left(t\right)
				\end{array}}}\left(x\left(t\right)-\hat{x}\left(t\right)\right) + \left.\frac{\partial f\left(x,p,u\right)}{\partial p}\right|_{\scriptsize {\begin{array}{l}
				x=\hat{x}\left(t\right)\\
				p=\hat{p}\\
				u=\hat{u}\left(t\right)
				\end{array}}}\left(p-\hat{p}\right)\nonumber +\left.\frac{\partial f\left(x,p,u\right)}{\partial u}\right|_{\scriptsize {\begin{array}{l}
				x=\hat{x}\left(t\right)\\
				p=\hat{p}\\
				u=\hat{u}\left(t\right)
				\end{array}}}\left(u\left(t\right)-\hat{u}\left(t\right)\right)\nonumber\\
		&= A\left(t\right)\delta_{x}\left(t\right)+B_{1}\left(t\right)\delta_{p}+B_{2}\left(t\right)\delta_{u}\left(t\right).\label{eq:LTV}
		\end{align}}
	From \cite{Athans1966}, for unconstrained $\delta_{u}\left(t\right)$, symmetric matrices $Q,S\succeq0$ and $R\succ0$, the optimal control of the stabilizable LTV system in (\ref{eq:LTV}) with $\delta_{x}\left(t\right)$ as output, and quadratic cost 
	{\footnotesize \[
		J_{LQR}=\frac{1}{2}\delta_{x}^{\top}\left(t_{f}\right)S\delta_{x}\left(t_{f}\right)+\frac{1}{2}\int_{t_{0}}^{t_{f}}\left( \delta_{x}^{\top}\left(t\right)Q\delta_{x}\left(t\right)+\delta_{u}^{\top}\left(t\right)R\delta_{u}\left(t\right)\right) \, dt
		\]}
	exists, is unique, time varying, and is given by 
	{\small \begin{align} \begin{split}
		\delta_{u}\left(t\right) & = -R^{-1}B_{2}^{\top}\left(t\right)P\left(t\right)\delta_{x}\left(t\right)\\
		& =: -K_{LQR}\left(t\right)\delta_{x}\left(t\right), \label{eq:LQRgain}
		\end{split} \end{align}}where, considering the boundary condition $P\left(t_{f}\right)=S$, $P\left(t\right)$ is the solution of the Riccati matrix differential equation 
	{\small \begin{equation}
		\dot{P}\left(t\right)=-P\left(t\right)A\left(t\right)-A^{\top}\left(t\right)P\left(t\right)+P\left(t\right)B_{2}\left(t\right)R^{-1}B_{2}^{\top}\left(t\right)P\left(t\right)-Q.\label{eq:Ric}
		\end{equation}} 
	The closed-loop nonlinear dynamics of the three-link robot modeling the PLLO and its user performing the STS movement under state feedback control become{\small \begin{align}
		\dot{x}\left(t\right) & =f\left(x\left(t\right),p,\hat{u}\left(t\right)-K_{LQR}\left(t\right)\left(x\left(t\right)-\hat{x}\left(t\right)\right)\right)\nonumber\\
		&=: f_{cl}\left(t,x\left(t\right),p\right). \label{eq:Nonlinear}
		\end{align}} 
	\section{SENSITIVITY-BASED REACHABILITY ANALYSIS UNDER PARAMETER UNCERTAINTY}
	
	In this section, we first review the method presented in~\cite{meyer2018sampled} to over-approximate the reachable sets of an uncertain dynamical system and then introduce an approach extending these results to auxiliary static systems, such as those defined by an output function or a feedback controller.
	For the sake of generality, we thus initially consider a time-varying system
	{\small \begin{align}
		\label{eq:system generic}
		\dot x(t) &=\varphi(t,x(t),p),
		\end{align}}where $x\in\R^{n_x}$ is the state and $p\in\R^{n_p}$ is a constant but uncertain parameter.
	Consider that \eqref{eq:system generic} has a single initial state $x_0\in\R^{n_x}$ at time $t_0\in\R$ and let $\underline{p},\overline{p}\in\R^{n_p}$ define the parameter uncertainty of \eqref{eq:system generic} as an interval $[\underline p,\overline p]\subseteq\R^{n_p}$.
	Then the trajectories of \eqref{eq:system generic} are denoted by function $\Phi$, where $\Phi(t;t_0,x_0,p)\in\R^{n_x}$ represents the successor reached at time $t\geq t_0$ by system \eqref{eq:system generic} starting from initial state $x_0$ and with constant parameter $p\in[\underline p,\overline p]$.
	Next, let {\small \begin{equation*}
		{\small RS(t,[\underline p,\overline p]):=\{\Phi(t;t_0,x_0,p)~|~p\in[\underline p,\overline p]\}\subseteq\R^{n_x}}
		\end{equation*}}denote the reachable set of \eqref{eq:system generic} at time $t\geq t_0$ for all possible parameter values in $[\underline p,\overline p]$, and {\small \begin{equation}
		\label{eq:sensitivity x-space}
		S(t;t_0,x_0,p):=\frac{\partial\Phi(t;t_0,x_0,p)}{\partial p}\in\R^{{n_x}\times {n_p}}
		\end{equation}}be the sensitivity of the trajectories of \eqref{eq:system generic} with respect to the parameter uncertainty.
	The reachability analysis in~\cite{meyer2018sampled} is based on a boundedness assumption on this sensitivity matrix at each time $t$.
	\begin{assum}
		\label{assum:bounded sensitivity}
		For all $(i,j)\in\{1,\dots,{n_x}\}\times\{1,\dots,{n_p}\}$ there exists $\underline{\mathcal{S}_{ij}}, \overline{\mathcal{S}_{ij}}:[t_0,+\infty)\rightarrow\R$ such that for all $t\geq t_0$ and $p\in [\underline p,\overline p]$ we have $S_{ij}(t;t_0,x_0,p)\in[\underline{\mathcal{S}_{ij}}(t),\overline{\mathcal{S}_{ij}}(t)]$.
	\end{assum}
	
	For each time $t\geq t_0$ and index $i\in\{1,\dots,{n_x}\}$, let parameter values $\underline{\pi}^i(t),\overline{\pi}^i(t)\in[\underline p,\overline p]$ and row vector $d^i(t)\in\R^{n_p}$ be written as follows{\small \begin{equation*}
		\begin{cases}
		\underline{\pi}^i(t):=[\underline{\pi}^i_1(t); \dots; \underline{\pi}^i_{n_p}(t)]\\
		\overline{\pi}^i(t):=[\overline{\pi}^i_1(t); \dots; \overline{\pi}^i_{n_p}(t)]\\
		d^i(t):=[d^i_1(t),\dots,d^i_{n_p}(t)]
		\end{cases}
		\end{equation*}}and whose elements are defined for each $j\in\{1,\dots,{n_p}\}$ based on the sign of the variable $\mathcal{S}_{ij}^*(t)$ denoting the center of the scalar interval $[\underline{\mathcal{S}_{ij}}(t),\overline{\mathcal{S}_{ij}}(t)]$:
	{\small \begin{equation}
		\label{eq:param bounds for OA}
		\begin{cases}
		\mathcal{S}^*_{ij}(t)\geq0\Rightarrow \underline{\pi}^i_j(t)=\underline{p}_j,\quad\overline{\pi}^i_j(t)=\overline{p}_j,\quad d^i_j(t)=\min(0,\underline{\mathcal{S}_{ij}}(t)),\\
		\mathcal{S}^*_{ij}(t)<0\Rightarrow \underline{\pi}^i_j(t)=\overline{p}_j,\quad\overline{\pi}^i_j(t)=\underline{p}_j,\quad d^i_j(t)=\max(0,\overline{\mathcal{S}_{ij}}(t)).\\
		\end{cases}
		\end{equation}}These vectors can then be used as in \cite{meyer2018sampled} to obtain over-approximations of the reachable sets of \eqref{eq:system generic}. 
	\begin{proposition} 
		\label{prop:over approximation}
		Under Assumption~\ref{assum:bounded sensitivity} and the definition of vectors $\underline{\pi}^i(t),\overline{\pi}^i(t),d^i(t)$ in \eqref{eq:param bounds for OA}, we define two functions $\underline{r},\overline{r}:[t_0,+\infty)\rightarrow\R^{n_x}$ such that for each $t\geq t_0$ and $i\in\{1,\dots,{n_x}\}$:
		{\small \begin{equation*}
			\begin{cases}
			\underline{r}_i(t)=\Phi_i(t;t_0,x_0,\underline{\pi}^i(t))-d^i(t)(\underline{\pi}^i(t)-\overline{\pi}^i(t)),\\
			\overline{r}_i(t)=\Phi_i(t;t_0,x_0,\overline{\pi}^i(t))+d^i(t)(\underline{\pi}^i(t)-\overline{\pi}^i(t)).
			\end{cases}  
			\end{equation*}}Then an interval over-approximation of the reachable set of \eqref{eq:system generic} at time $t$ is given by $RS(t,[\underline p,\overline p])\subseteq[\underline{r}(t),\overline{r}(t)].$
		If in addition $d^i_j(t)=0$ for all $(i,j)\in\{1,\dots,{n_x}\}\times\{1,\dots,{n_p}\}$, then $[\underline{r}(t),\overline{r}(t)]$ is a tight over-approximation (smallest interval containing $RS(t,[\underline p,\overline p])$).
	\end{proposition}
	
	The result presented in Proposition~\ref{prop:over approximation} is applicable to any system described by a trajectory function $\Phi$.
	In the remainder of this section, we aim to apply this approach not only to the dynamical system \eqref{eq:system generic}, but also to two auxiliary static systems to be defined later.
	To distinguish these systems, we thus denote with the superscript $x$ (e.g.\ $\Phi^x$, $S^x$, $RS^x$) the variable specifically related to the dynamical system \eqref{eq:system generic}.
	
	In order to use Proposition~\ref{prop:over approximation} on system \eqref{eq:system generic} we first need to obtain bounds on its sensitivity matrix at each time $t$ as in Assumption~\ref{assum:bounded sensitivity}.
	For this, we first apply the chain rule to the sensitivity definition \eqref{eq:sensitivity x-space} to obtain a time-varying affine system that describes the evolution of the sensitivity matrix~\cite{khalil2001nonlinear} in terms of the Jacobian matrices of \eqref{eq:Nonlinear} evaluated along the trajectory $\Phi^x(t;t_0,x_0,p)$: {\small \begin{align}
		\dot S^x(t;t_0,x_0,p)=&\left.\frac{\partial f_{cl}(t,x,p)}{\partial x}\right|_{x=\Phi^x(t;t_0,x_0,p)} S^x(t;t_0,x_0,p) + \left.\frac{\partial f_{cl}(t,x,p)}{\partial p}\right|_{x=\Phi^x(t;t_0,x_0,p)},\label{eq:sensitivity system}
		\end{align}}which is initialized with the zero matrix $S^x(t_0;t_0,x_0,p)=0_{n_x\times n_p}$.
	
	The sensitivity bounds $[\underline{\mathcal{S}^x_{ij}}(t),\overline{\mathcal{S}^x_{ij}}(t)]$ for \eqref{eq:system generic} at time $t\geq t_0$ can then be estimated through a sampling approach consisting in first solving the sensitivity system \eqref{eq:sensitivity system} numerically over $[t_{0},t_{f}]$ for a finite set of randomly chosen parameters $\mathcal{P}\subset[\underline{p},\overline{p}]$.
	Then, for each time $t\in[t_{0},t_{f}]$ where $\underline{r}(t)$ and $\overline{r}(t)$ are to be computed, and each element $S^x_{ij}$ of the sensitivity matrix \eqref{eq:sensitivity x-space}, an approximation $[\underline{{S}^x_{ij}}(t),\overline{{S}^x_{ij}}(t)]$ of the bounds $[\underline{\mathcal{S}^x_{ij}}(t),\overline{\mathcal{S}^x_{ij}}(t)]$ in Assumption~\ref{assum:bounded sensitivity} is obtained from the extremal values of the computed sensitivities over the set of parameter samples $\mathcal{P}$:{\small \begin{equation}
		\begin{cases}
		\displaystyle\overline{S^x_{ij}}(t)=\max_{p\in\mathcal{P}}S^x_{ij}(t;t_0,x_0,p),\\
		\displaystyle\underline{S^x_{ij}}(t)=\min_{p\in\mathcal{P}}S^x_{ij}(t;t_0,x_0,p).
		\end{cases} \label{eq:SamplingApproach}
		\end{equation}}
	Since the resulting bounds $[\underline{S^x_{ij}}(t),\overline{S^x_{ij}}(t)]$ are not guaranteed to satisfy Assumption \ref{assum:bounded sensitivity}, a more reliable approximation may be found by iteratively enlarging these bounds through a falsification approach. An iteration of the falsification at time $t$ looks for parameters in $[\underline p,\overline p]$ whose sensitivity does not lie within the bounds from the sampling approach, which is achieved by solving the optimization problem {\footnotesize \begin{equation*}
		J_{F}\left(t\right):=\min_{p\in[\underline p,\overline p]}\left(\min_{i,j}\left(\frac{\overline{S^x_{ij}}(t)-\underline{S^x_{ij}}(t)}{2}-\left|S^x_{ij}(t;t_0,x_0,p)-
		\frac{\underline{S^x_{ij}}(t)+\overline{S^x_{ij}}(t)}{2}
		\right|\right)\right).
		\end{equation*}}The cost function used in this minimization problem is defined for each pair $(i,j)\in\{1,\dots,n_x\}\times\{1,\dots,n_p\}$ by an inverted and translated absolute value function such that it returns a negative value if and only if $S^x_{ij}(t;t_0,x_0,p)\notin[\underline{S^x_{ij}}(t),\overline{S^x_{ij}}(t)]$.
	As a result, finding $J_F(t)<0$ guarantees that there exists a pair $(i,j)\in\{1,\dots,n_x\}\times\{1,\dots,n_p\}$ for which the sensitivity bounds $[\underline{S^x_{ij}}(t),\overline{S^x_{ij}}(t)]$ have been falsified.
	These bounds thus need to be updated according to the sensitivity value $S_{ij}^x(t;t_0,x_0,p^*)$ for the optimizer $p^*\in[\underline p,\overline p]$ associated with the obtained local minimum.
	This falsification procedure is then repeated until we obtain $J_F(t)\geq 0$.
	
	\begin{remark}
		\label{rmk:falsification}
		Although falsification can help to improve the approximation of the sensitivity bounds, it cannot provide formal guarantees that Assumption~\ref{assum:bounded sensitivity} is satisfied with the enlarged bounds, because the optimization problem can only find local minima. There exists an alternative approach based on interval analysis presented in~\cite{meyer2018sampled} for which such guarantees are provided, but it has been shown to be of limited practical use, due to the overly conservative nature of the obtained sensitivity bounds.
	\end{remark}
	
	Applying Proposition~\ref{prop:over approximation} with the obtained sensitivity bounds $\underline{S^x},\overline{S^x}:[t_0,+\infty)\rightarrow\R^{n_x\times n_p}$ thus results in two functions $\underline{r^x},\overline{r^x}:[t_0,+\infty)\rightarrow\R^{n_x}$ over-approximating the reachable set of \eqref{eq:system generic} at each time $t\geq t_0$:{\small \begin{equation}
		\label{eq:over-approx x}
		RS^x(t,[\underline p,\overline p]):=\{\Phi^x(t;t_0,x_0,p)~|~p\in[\underline p,\overline p]\}\subseteq[\underline{r^x}(t),\overline{r^x}(t)].
		\end{equation}}
	
	Consider now an output map $\zeta:\R^{n_x}\times[\underline p,\overline p]\rightarrow\R^{n_y}$ defining the output $y=\zeta(x,p)$ of system \eqref{eq:system generic} based on its state and parameter.
	A reachability analysis on the output $y\in\R^{n_y}$ is thus done by applying Proposition~\ref{prop:over approximation} to the static system describing the evolution of $y$ in terms of the trajectories of $x$:{\small \begin{equation}
		\label{eq:static system y}
		\Psi^y(t;t_0,x_0,p):=\zeta(\Phi^x(t;t_0,x_0,p),p).
		\end{equation}}Similarly to \eqref{eq:sensitivity x-space}, we can define the sensitivity $S^y$ of \eqref{eq:static system y} with respect to the parameter $p$ and then use the chain rule on $\zeta$ to relate it to $S^x$:{\small \begin{align}
		S^{y}\left(t;t_0,x_0,p\right)&:=\frac{\partial\Psi^{y}\left(t;t_0,x_0,p\right)}{\partial p}=\frac{\partial}{\partial p}\left(\zeta\left(\Phi^{x}\left(t;t_0,x_0,p\right),p\right)\right)\nonumber\\
		&=\left.\frac{\partial\zeta\left(x,p\right)}{\partial x}\right|_{x=\Phi^{x}\left(t;t_0,x_0,p\right)}\frac{\partial\Phi^{x}\left(t;t_0,x_0,p\right)}{\partial p} + \left.\frac{\partial\zeta\left(x,p\right)}{\partial p}\right|_{x=\Phi^{x}\left(t;t_0,x_0,p\right)}\nonumber\\
		&=\left.\frac{\partial\zeta\left(x,p\right)}{\partial x}\right|_{x=\Phi^{x}\left(t;t_0,x_0,p\right)}S^{x}\left(t;t_0,x_0,p\right)+\left.\frac{\partial\zeta\left(x,p\right)}{\partial p}\right|_{x=\Phi^{x}\left(t;t_0,x_0,p\right)} \label{eq:sensitivity y}
		\end{align}}
	
	With knowledge of the sensitivity bounds $\underline{S^x},\overline{S^x}:[t_0,+\infty)\rightarrow\R^{n_x\times n_p}$ for \eqref{eq:system generic} and the mapping $\zeta:\R^{n_x}\times[\underline p,\overline p]\rightarrow\R^{n_y}$, the sensitivity bounds $\underline{S^y},\overline{S^y}:[t_0,+\infty)\rightarrow\R^{n_y\times n_p}$ for the static system \eqref{eq:static system y} can be computed. Equation \eqref{eq:param bounds for OA} is thus reused with $\underline{S^y},\overline{S^y}$ to apply Proposition~\ref{prop:over approximation} on \eqref{eq:static system y} and obtain over-approximation functions $\underline{r^y},\overline{r^y}:[t_0,+\infty)\rightarrow\R^{n_y}$ such that for each time $t\geq t_0$:{\small \begin{equation}
		\label{eq:over-approx y}
		RS^y(t,[\underline p,\overline p]):=\{\Psi^y(t;t_0,x_0,p)~|~p\in[\underline p,\overline p]\}\subseteq[\underline{r^y}(t),\overline{r^y}(t)].
		\end{equation}}
	
	Assuming that \eqref{eq:system generic} is actually a closed-loop system obtained from the use of a feedback controller $u(t)=K(t,x(t),p)$ with $K:[t_0,+\infty)\times\R^{n_x}\times\R^{n_p}\rightarrow\R^{n_u}$, we can apply the same approach as for the output $y$ by defining the static system{\small \begin{equation}
		\label{eq:static system u}
		\Psi^u(t;t_0,x_0,p):=K(t,\Phi^x(t;t_0,x_0,p),p).
		\end{equation}}The sensitivity $S^u$ of \eqref{eq:static system u} with respect to the parameter $p$ is then obtained similarly to $S^y$ in \eqref{eq:sensitivity y}:
	{\small \begin{align}
		S^{u}\left(t;t_0,x_0,p\right)&:=\frac{\partial\Psi^{u}\left(t;t_0,x_0,p\right)}{\partial p} = \frac{\partial}{\partial p}(K(t,\Phi^x(t;t_0,x_0,p),p))\nonumber\\
		&=\left.\frac{\partial K\left(t,x,p\right)}{\partial x}\right|_{x=\Phi^{x}\left(t;t_0,x_0,p\right)}S^{x}\left(t;t_0,x_0,p\right) + \left.\frac{\partial K\left(t,x,p\right)}{\partial p}\right|_{x=\Phi^{x}\left(t;t_0,x_0,p\right)}\label{eq:sensitivity u}
		\end{align}}which then leads to sensitivity bounds $\underline{S^u},\overline{S^u}:[t_0,+\infty)\rightarrow\R^{n_u\times n_p}$ for the static system \eqref{eq:static system u} to be used in Proposition~\ref{prop:over approximation} and obtain over-approximation functions $\underline{r^u},\overline{r^u}:[t_0,+\infty)\rightarrow\R^{n_u}$ such that for each time $t\geq t_0$ we have {\small \begin{equation}
		\label{eq:over-approx u}
		RS^u(t,[\underline p,\overline p]):=\{\Psi^u(t;t_0,x_0,p)~|~p\in[\underline p,\overline p]\}\subseteq[\underline{r^u}(t),\overline{r^u}(t)].
		\end{equation}}
	\section{NUMERICAL APPLICATION OF THE REACHABILITY ANALYSIS FOR AN STS MOVEMENT}
	
	The ascending phase of the STS movement under study starts from rest, with the shank and torso segments parallel to the vertical, and the thigh segment parallel to the horizontal, by setting $x\left(t_{0}\right)=\left[90\text{\textdegree};-90\text{\textdegree};90\text{\textdegree};0;0;0 \right]$. With nominal parameter values {\small \begin{align*}
		\hat{p} &= \left[ 9.68; \;\; 12.59; \;\; 44.57; \;\; 1.16; \;\; 0.52; \;\; 2.56; \;\; 0.53; \;\; 0.41; \;\; 0.52; \;\; 0.265; \;\; 0.205; \;\; 0.26 \right],
		\end{align*}}the corresponding initial position of the CoM of the three-link robot is $(\hat{x}_{CoM}\left(t_{0}\right),\hat{y}_{CoM}\left(t_{0}\right))=(0.309,0.6678)\left[m\right]$.
	
	For planning the rest-to-rest maneuver from $ \hat{z}\left(t\right) $, $\dot{\hat{z}}\left(t\right)$ and $\ddot{\hat{z}}\left(t\right)$ in (\ref{eq:MotionPlanning}), define {\small $\Theta_{i}\left(t,t_{f}\right):=-2\frac{t^{3}}{t_{f}^{3}}+3\frac{t^{2}}{t_{f}^{2}}$} for $i=1,2,3$, which is the only cubic polynomial satisfying {\small $\dot{\Theta}_{i}\left(t_{0},t_{f}\right)=\dot{\Theta}_{i}\left(t_{f},t_{f}\right)=0$}, {\small $\Theta_{i}\left(t_{0},t_{f}\right)=0$}, and {\small $\Theta_{i}\left(t_{f},t_{f}\right)=1$}. 
	Considering $t_{0}=0$ and $t_{f}=3.5\left[s\right]$ and a final configuration that places the CoM directly above the origin of the inertial frame with the values $\hat{\theta}_{2}\left(t_{f}\right)=-5\text{\textdegree}$, $\hat{x}_{CoM}\left(t_{f}\right)=0$ and $\hat{y}_{CoM}\left(t_{f}\right)=0.974\left[m\right]$, the reference state trajectories $\hat{x}\left(t\right)$ can be determined from (\ref{eq:z2theta}).
	
	When solving for $ \hat{u}\left(t\right) $ in (\ref{eq:Allocation}), it is enforced that the contributions from $\tau_{h}\left(t\right)$, $\tau_{s}\left(t\right)$ and $F_{y}\left(t\right)$ outweigh $F_{x}\left(t\right)$ taking $W_u=\operatorname{diag}\left(\left[ 1, 1, 10, 1\right]\right)$ and, because the user of the PLLO always pushes the crutches down to propel upwards, the constraint $F_{y}\left(t\right)\geq0$ is imposed; all other inputs are unconstrained. After numerically computing the linearization in \eqref{eq:LTV}, the weight matrices from \cite{Narvaez-Aroche2018} 
	{\small \begin{align*}
		Q & = \operatorname{diag}\left(\left[3237, \;5534, \;6546, \;7918, \;4003, \;8516\right]\right)\\
		R & = \operatorname{diag}\left(\left[0.3659, \;0.0155, \;0.1433, \;0.1553\right]\right)\\
		S & = \operatorname{diag}\left(\left[1068, \;5396, \;1324, \;9467, \;3975, \;5819\right]\right)
		\end{align*}}are plugged into \eqref{eq:Ric}, which is solved with tools documented in \cite{Moore2015} to obtain their corresponding time-varying gain $ K_{LQR} \left( t \right) \in \mathbb{R}^{4 \times 6} $ from~\eqref{eq:LQRgain}. Using this gain for the state feedback control of the STS movement, brings the dynamics of the three-link robot modeling the PLLO and its user to the closed-loop form in \eqref{eq:Nonlinear}.

	Considering a sampling of the time horizon $[0,3.5]$ at a frequency of $100\,[ Hz ]$ to obtain a set of $351$ sampled times denoted as $T_s:= \left\{ 0:0.01:3.5 \right\}$; the goal of this section is to apply the sensitivity-based reachability analysis to compute, at each time $t\in T_s$, the over-approximations $[\underline{r^x}(t),\overline{r^x}(t)]$, $[\underline{r^y}(t),\overline{r^y}(t)]$ and $[\underline{r^u}(t),\overline{r^u}(t)]$ defined in \eqref{eq:over-approx x}, \eqref{eq:over-approx y} and \eqref{eq:over-approx u}, for the state $x=[\theta_1;\theta_2;\theta_3;\dot\theta_1;\dot\theta_2;\dot\theta_3]$, the output defined as $y:=[x_{CoM};y_{CoM};\dot x_{CoM};\dot y_{CoM}]$ and the control input $u=[\tau_{h};\tau_{s};F_{x};F_{y}]$.
	The parameter uncertainties lie within the interval $[\underline p,\overline p]\subseteq\R^{12}$ in Table~\ref{tab:Nominal}, which was calculated for a fluctuation of $\pm5\%$ of the nominal weight of the user with anthropometric data from~\cite{Bartel2006}. 
	\begin{table}[h]
		\caption{Bounds for the Parameter Uncertainties of the System $ [\underline{p}, \overline{p} ]$ \label{tab:Nominal}}
		\centering{}
		\begin{tabular}{ccccc}
			\toprule
			Link & $m_{i}\:\left[kg\right]$ & $I_{i}\:\left[kg\cdot m^{2}\right]$ & $l_{i}\:\left[m\right]$ & $l_{ci}\:\left[m\right]$\tabularnewline
			\midrule
			\midrule 
			1 & $\left[9.2, 10.2\right]$ & $\left[1.10, 1.21\right]$ & $\left[0.52, 0.54\right]$ & $\left[0.23, 0.30\right]$\tabularnewline
			\midrule 
			2 & $\left[11.2, 13.2\right]$ & $\left[0.49, 0.54\right]$ & $\left[0.39, 0.42\right]$ & $\left[0.17, 0.23\right]$\tabularnewline
			\midrule 
			3 & $\left[42.3, 46.8\right]$ & $\left[2.40, 2.65\right]$ & $\left[0.51, 0.53\right]$ & $\left[0.24, 0.28\right]$\tabularnewline
			\bottomrule
		\end{tabular}
	\end{table}

	With a set $\mathcal{P}_b\subset[\underline p,\overline p]$ of $500$ parameters drawn from a Latin Hypercube, the first step in the analysis is to numerically solve the sensitivity equation \eqref{eq:sensitivity system} over the time horizon $[0,3.5]$ for all $p \in \mathcal{P}_b$. According to the sampling approach of the previous section, the sensitivity bounds $\underline{S^x},\overline{S^x}:[0,3.5]\rightarrow\R^{6\times 12}$ are then estimated by minimizing/maximizing the entries of the matrices~\eqref{eq:sensitivity x-space} for each $t \in T_s$  as in \eqref{eq:SamplingApproach}.
	
\begin{figure}[h]
		\begin{centering}
			\includegraphics[width=7.5cm]{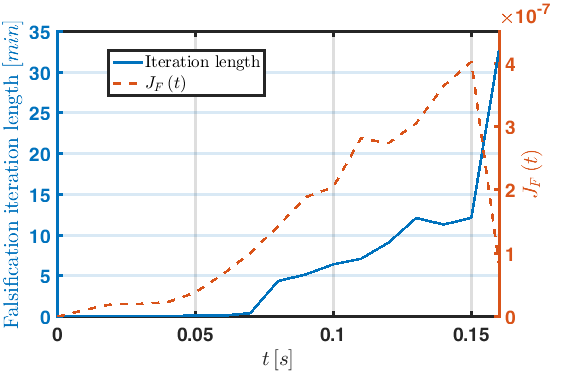}
			\par\end{centering}
		\caption{Length of a falsification iteration at time $t$ and cost {\small $J_{F}\left(t\right)$}. \label{fig:FalsificationMetrics} }
	\end{figure}

	The sensitivity bounds from sampling may be refined through the falsification approach presented in the previous section. The time spent in a single falsification iteration over the bounds estimated by sampling $\underline{S^x}\left(t\right),\overline{S^x}\left(t\right)$ for the first 17 elements in $T_s$ are shown in Figure~\ref{fig:FalsificationMetrics}, together with the calculated cost $J_{F} \left( t \right)$. It can be seen that the falsification is done quickly for the first few time steps, but going further into $T_s$, it grows to the point where it becomes unpractical to continue executing it. In addition, the positive values of $J_{F} \left( t \right)$ mean that the first iteration of the falsification does not provide any improvement of the sensitivity obtained in the sampling approach. On the basis of these observations and of Remark~\ref{rmk:falsification}, the results that follow only rely on the sampling approach with the assumption that the sensitivity bounds $[\underline{S^x_{ij}}(t),\overline{S^x_{ij}}(t)]$ obtained from the exploration of the solutions of the sensitivity equation $\forall p \in \mathcal{P}_b$ are close enough to an over-approximation of the set $\{S^x_{ij}(t;t_0,x_0,p)~|~p\in[\underline{p},\overline{p}]\}$, as required in Assumption~\ref{assum:bounded sensitivity}. A consequence of this assumption is that even though the reachability analysis result in Proposition~\ref{prop:over approximation} might not always be a true over-approximation of the reachable set, it still provides an accurate measure of the worst-case performances for the closed-loop system \eqref{eq:Nonlinear}.

\begin{figure}[H]
	\centering
	\subfloat[Angular position of link 1 relative to the horizontal.\label{fig:theta1} ]{\includegraphics[width=7.5cm]{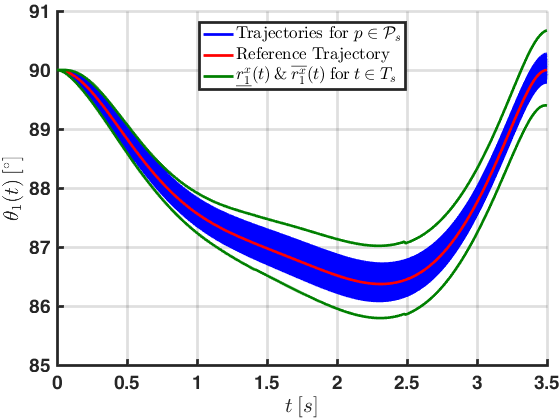}} \quad \quad \quad
	\subfloat[Angular position of link 2 relative to link 1.\label{fig:theta2} ]{\includegraphics[width=7.5cm]{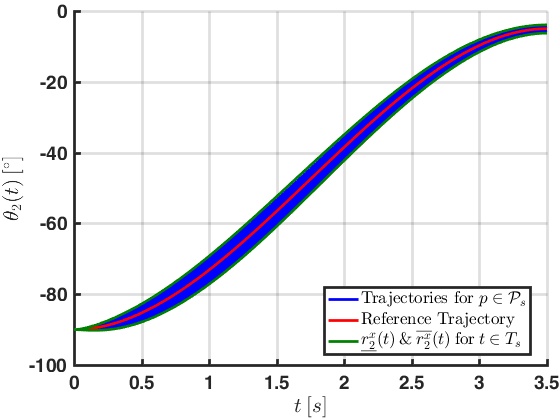}}
	\\
	\subfloat[Angular position of link 3 relative to link 2.\label{fig:theta3}]{ \includegraphics[width=7.5cm]{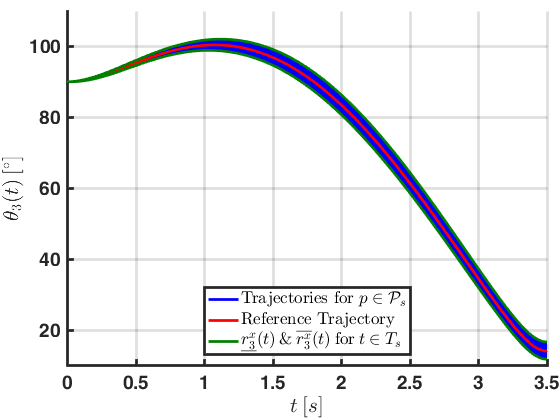}} \quad \quad \quad
	\subfloat[Angular velocity of link 1.\label{fig:omega1}]{ \includegraphics[width=7.5cm]{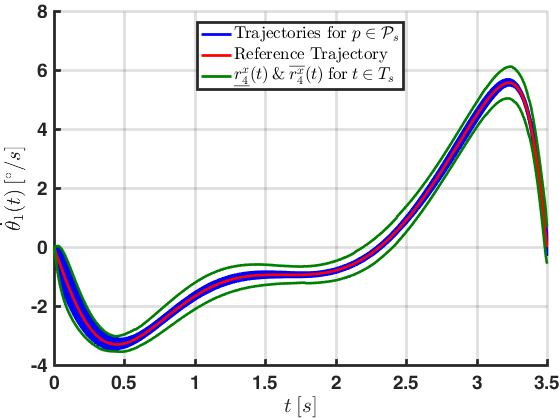}}
	\\
	\subfloat[Angular velocity of link 2.\label{fig:omega2}]{ \includegraphics[width=7.5cm]{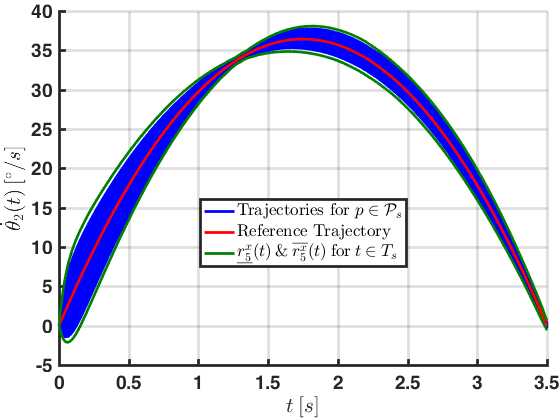}} \quad \quad \quad
	\subfloat[Angular velocity of link 3.\label{fig:omega3}]{ \includegraphics[width=7.5cm]{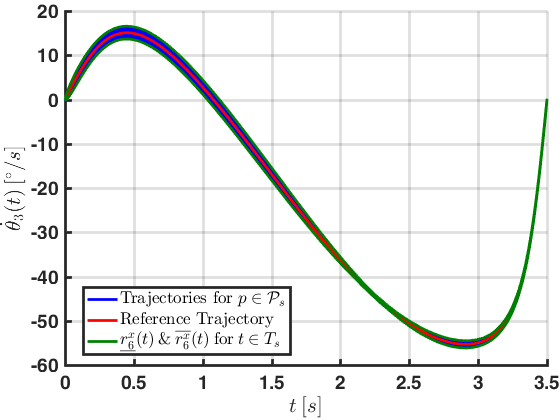}}
	\caption{State over-approximations $ [\underline{r^x}(t),\overline{r^x}(t)] $ for every $t\in T_{s}$ during the STS movement.\label{fig:State}}
\end{figure}

	Once $[\underline{S^x_{ij}}(t),\overline{S^x_{ij}}(t)]$ are known, Proposition~\ref{prop:over approximation} is applied to obtain the over-approximations $[\underline{r^x}(t),\overline{r^x}(t)]$ for every $t\in T_{s}$, which are displayed in green in Figure~\ref{fig:State} for each state in $x$. To visualize their tightness, the plots also provide, in blue, the trajectories of the closed-loop system \eqref{eq:Nonlinear} for a set $\mathcal{P}_{s}\subseteq[\underline p,\overline p]$ of $500$ parameters from a Latin Hypercube sampling (note that this set is different from $\mathcal{P}_{b}$). The reference trajectory $\Phi^x(t;0,x_0,\hat p)$ of \eqref{eq:Nonlinear} for $\hat p$ is in red. 
	The over-approximations for $\theta_{1}\left(t\right)$ in Figure~\ref{fig:theta1} show that the terminal position of the shank segment under the parameter uncertainties will only be slightly off the vertical ($\pm 0.5\text{\textdegree}$), easing the stabilization phase for completing standing. The ones for $\theta_{2}\left(t\right)$ in Figure~\ref{fig:theta2} do not become positive, meaning that the controller will not cause the knee of the user to hyperextend. Also, since $\theta_{3}\left(t\right)$ in Figure~\ref{fig:theta3} never goes negative and only approaches zero at the end of the horizon, the torso will have natural configurations while ascending.

\begin{figure}[H]
		\centering
		\subfloat[$x_{CoM}$ coordinate of the position of the three-link robot CoM.\label{fig:xCoM}]{\includegraphics[width=7.5cm]{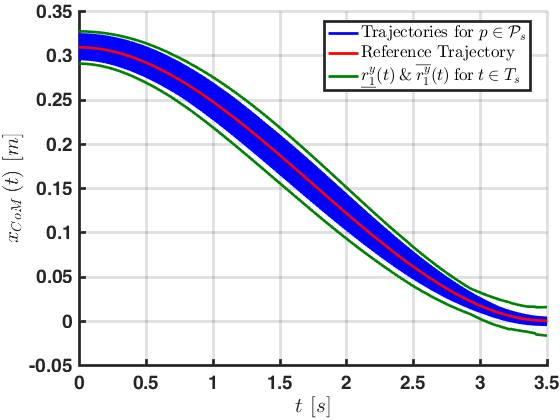}} \quad \quad \quad
		\subfloat[$y_{CoM}$ coordinate of the position of the three-link robot CoM.\label{fig:yCoM}]{\includegraphics[width=7.5cm]{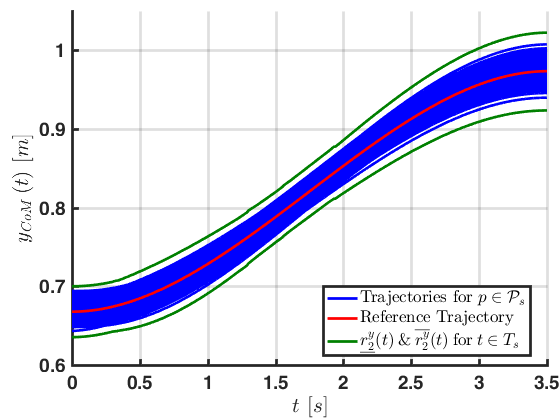}}
		\\
		\subfloat[$\dot{x}_{CoM}$ coordinate of the velocity of the three-link robot CoM.\label{fig:vxCoM}]{ \includegraphics[width=7.5cm]{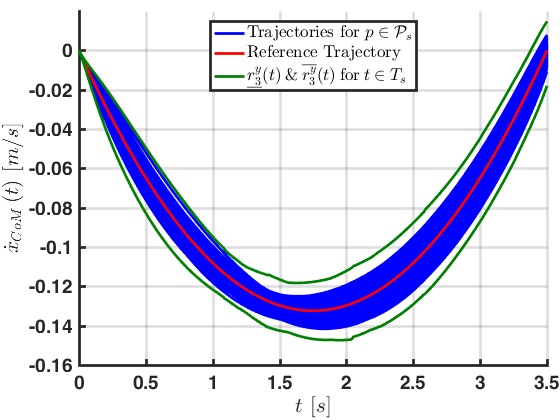}} \quad \quad \quad
		\subfloat[$\dot{y}_{CoM}$ coordinate of the velocity of the three-link robot CoM.\label{fig:vyCoM}]{ \includegraphics[width=7.5cm]{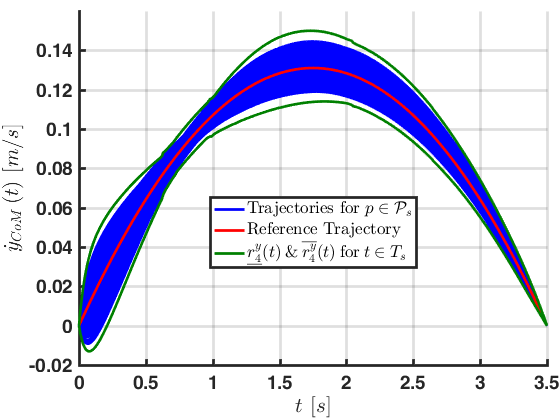}}
		\caption{Output over-approximations $ [\underline{r^y}(t),\overline{r^y}(t)] $ for every $t\in T_{s}$ during the STS movement.\label{fig:Output}}
	\end{figure}

	
	The output~$y=[x_{CoM};y_{CoM};\dot x_{CoM};\dot y_{CoM}]$ is computed with the mapping $\zeta:\R^6\times[\underline p,\overline p]\rightarrow\R^4$ defined from the kinematic equations for the CoM of the three-link planar robot in Figure~\ref{fig:Robot} that were derived in~\cite{Narvaez-Aroche2017}:{\small \begin{align}
		y=\left[\begin{array}{c}
		x_{CoM}\\ y_{CoM}\\ \dot{x}_{CoM}\\ \dot{y}_{CoM}
		\end{array}\right] & = \left[\begin{array}{c}
		k_{0}\left(k_{1}c_{1}+k_{2}c_{12}+k_{3}c_{123}\right)\\
		k_{0}\left(k_{1}s_{1}+k_{2}s_{12}+k_{3}s_{123}\right)\\
		-\dot{\theta}_{1}y_{CoM}-\dot{\theta}_{2}k_{0}\left(k_{2}s_{12}+k_{3}s_{123}\right)-\dot{\theta}_{3}k_{0}k_{3}s_{123}\\
		\dot{\theta}_{1}x_{CoM}+\dot{\theta}_{2}k_{0}\left(k_{2}c_{12}+k_{3}c_{123}\right)+\dot{\theta}_{3}k_{0}k_{3}c_{123}
		\end{array}\right]\nonumber\\
		& =: \zeta\left(x,p\right).\label{eq:zeta}
		\end{align}}
	Defining {\small \begin{align*}
		k_{4}\left(\theta,p\right) &:=k_{0}\left(p\right)\left(k_{2}\left(p\right)s_{12}+k_{3}\left(p\right)s_{123}\right),\\
		k_{5}\left(\theta,p\right) &:=k_{0}\left(p\right)k_{3}\left(p\right)s_{123},\\
		k_{6}\left(\theta,p\right) &:=k_{0}\left(p\right)\left(k_{2}\left(p\right)c_{12}+k_{3}\left(p\right)c_{123}\right), \\ 
		k_{7}\left(\theta,p\right) &:=k_{0}\left(p\right)k_{3}\left(p\right)c_{123},\\
		k_{8}\left(x,p\right) &:=k_{0}\left(p\right)\left(\left(k_{2}\left(p\right)s_{12}+k_{3}\left(p\right)s_{123}\right)\dot{\theta}_{2}+l_{c3}m_{3}s_{123}\dot{\theta}_{3}\right),\\
		k_{9}\left(x,p\right) &:=k_{0}\left(p\right)\left(\left(k_{2}\left(p\right)c_{12}+k_{3}\left(p\right)c_{123}\right)\dot{\theta}_{2}+l_{c3}m_{3}c_{123}\dot{\theta}_{3}\right),
		\end{align*}}the partial derivative of~\eqref{eq:zeta} with respect to $x$ is written as {\small \begin{align}
		\frac{\partial\zeta\left(x,p\right)}{\partial x}=
		\left[\begin{array}{cc}
		\zeta_{11}^{x} & 0\\
		\zeta_{21}^{x} & \zeta_{11}^{x}
		\end{array}\right]\in\R^{4\times 6}, \label{eq:dzetadx}
		\end{align}}with entries $\zeta_{ij}^{x}\in\R^{2\times 3}$ given by {\small \begin{align*}
		\zeta_{11}^{x}&=\left[\begin{array}{ccc}
		-y_{CoM} & -k_{4}\left(\theta,p\right) & -k_{5}\left(\theta,p\right)\\
		x_{CoM} & k_{6}\left(\theta,p\right) & k_{7}\left(\theta,p\right)
		\end{array}\right],\\
		\zeta_{21}^{x}&=-\left[\begin{array}{ccc}
		x_{CoM}\dot{\theta}_{1}+k_{6}\left(\theta,p\right)\dot{\theta}_{2}+k_{7}\left(\theta,p\right)\dot{\theta}_{3} & k_{7}\left(\theta,p\right)\dot{\theta}_{3} & 0\\
		y_{CoM}\dot{\theta}_{1}+k_{4}\left(\theta,p\right)\dot{\theta}_{2}+k_{5}\left(\theta,p\right)\dot{\theta}_{3} & k_{5}\left(\theta,p\right)\dot{\theta}_{3} & 0
		\end{array}\right]-\left[\begin{array}{ccc}
		0 & k_{6}\left(\theta,p\right)\left(\dot{\theta}_{1}+\dot{\theta}_{2}\right) & k_{7}\left(\theta,p\right)\left(\dot{\theta}_{1}+\dot{\theta}_{2}+\dot{\theta}_{3}\right)\\
		0 & k_{4}\left(\theta,p\right)\left(\dot{\theta}_{1}+\dot{\theta}_{2}\right) & k_{5}\left(\theta,p\right)\left(\dot{\theta}_{1}+\dot{\theta}_{2}+\dot{\theta}_{3}\right)
		\end{array}\right].
		\end{align*}}
\pagebreak
\pagebreak
	The partial derivative of \eqref{eq:zeta} with respect to $p$ is{\small \begin{align}
		\frac{\partial\zeta\left(x,p\right)}{\partial p}=\left[\begin{array}{cccc}
		\zeta_{11}^{p} & 0 & \zeta_{13}^{p} & \zeta_{14}^{p}\\
		\zeta_{21}^{p} & 0 & \zeta_{23}^{p} & \zeta_{24}^{p}
		\end{array}\right]\in\R^{4\times 12},\label{eq:dzetadp}
		\end{align}}where the entries $\zeta_{ij}^{p}\in\R^{2\times 3}$ are{\small \begin{align*}
		\zeta_{11}^{p}&=k_{0}\left(p\right)\left[\begin{array}{ccc}
		l_{c1}c_{1} & l_{1}c_{1}+l_{c2}c_{12} & l_{1}c_{1}+l_{2}c_{12}+l_{c3}c_{123}\\
		l_{c1}s_{1} & l_{1}s_{1}+l_{c2}s_{12} & l_{1}s_{1}+l_{2}s_{12}+l_{c3}s_{123}
		\end{array}\right] -k_{0}\left(p\right)\left[\begin{array}{ccc}
		x_{CoM} & x_{CoM} & x_{CoM}\\
		y_{CoM} & y_{CoM} & y_{CoM}
		\end{array}\right],\\
		\zeta_{13}^{p}&=k_{0}\left(p\right)\left[\begin{array}{ccc}
		\left(m_{2}+m_{3}\right)c_{1} & m_{3}c_{12} & 0\\
		\left(m_{2}+m_{3}\right)s_{1} & m_{3}s_{12} & 0
		\end{array}\right],\\
		\zeta_{14}^{p}&=k_{0}\left(p\right)\left[\begin{array}{ccc}
		m_{1}c_{1} & m_{2}c_{12} & m_{3}c_{123}\\
		m_{1}s_{1} & m_{2}s_{12} & m_{3}s_{123}
		\end{array}\right],\\
		\zeta_{21}^{p}&=k_{0}\left(p\right)\dot{\theta}_{1}\left[\begin{array}{ccc}
		-l_{c1}s_{1} & -\left(l_{1}s_{1}+l_{c2}s_{12}\right) & -\left(l_{1}s_{1}+l_{2}s_{12}+l_{c3}s_{123}\right)\\
		l_{c1}c_{1} & l_{1}c_{1}+l_{c2}c_{12} & l_{1}c_{1}+l_{2}c_{12}+l_{c3}c_{123}
		\end{array}\right] +k_{0}\left(p\right)\dot{\theta}_{1}\left[\begin{array}{ccc}
		y_{CoM} & y_{CoM} & y_{CoM}\\
		-x_{CoM} & -x_{CoM} & -x_{CoM}
		\end{array}\right]\\&\quad+k_{0}\left(p\right)\dot{\theta}_{2}\left[\begin{array}{ccc}
		0 & -l_{c2}s_{12} & -\left(l_{2}s_{12}+l_{c3}s_{123}\right)\\
		0 & l_{c2}c_{12} & \left(l_{2}c_{12}+l_{c3}c_{123}\right)
		\end{array}\right]+k_{0}\left(p\right)\left[\begin{array}{ccc}
		k_{8}\left(x,p\right) & k_{8}\left(x,p\right) & k_{8}\left(x,p\right)-l_{c3}s_{123}\dot{\theta}_{3}\\
		-k_{9}\left(x,p\right) & -k_{9}\left(x,p\right) & l_{c3}c_{123}\dot{\theta}_{3}-k_{9}\left(x,p\right)
		\end{array}\right],\\
		\zeta_{23}^{p}&=k_{0}\left(p\right)\left[\begin{array}{ccc}
		-\left(m_{2}+m_{3}\right)s_{1}\dot{\theta}_{1} & -m_{3}s_{12}\left(\dot{\theta}_{1}+\dot{\theta}_{2}\right) & 0\\
		\left(m_{2}+m_{3}\right)c_{1}\dot{\theta}_{1} & m_{3}c_{12}\left(\dot{\theta}_{1}+\dot{\theta}_{2}\right) & 0
		\end{array}\right],\\
		\zeta_{24}^{p}&=k_{0}\left(p\right)\left[\begin{array}{ccc}
		-m_{1}s_{1}\dot{\theta}_{1} & -m_{2}s_{12}\left(\dot{\theta}_{1}+\dot{\theta}_{2}\right) & -m_{3}s_{123}\left(\dot{\theta}_{1}+\dot{\theta}_{2}+\dot{\theta}_{3}\right)\\
		m_{1}c_{1}\dot{\theta}_{1} & m_{2}c_{12}\left(\dot{\theta}_{1}+\dot{\theta}_{2}\right) & m_{3}c_{123}\left(\dot{\theta}_{1}+\dot{\theta}_{2}+\dot{\theta}_{3}\right)
		\end{array}\right].
		\end{align*}}Plugging \eqref{eq:dzetadx} and \eqref{eq:dzetadp} into \eqref{eq:sensitivity y} we obtain the sensitivity bounds $\underline{S^y},\overline{S^y}:[0,3.5]\rightarrow\R^{6\times 12}$ and use Proposition~\ref{prop:over approximation} on \eqref{eq:static system y} to calculate the over-approximation bounds $\underline{r^y}(t),\overline{r^y}(t)$ for $t\in T_s$, shown in Figures~\ref{fig:xCoM}--\ref{fig:vyCoM} in green. The reference trajectory $\zeta(\Phi^x(t;0,x_0,\hat p),\hat p)$ is in red, and the trajectories $\zeta(\Phi^x(t;0,x_0,p),p)$ for all $p\in \mathcal{P}_s$ are in blue. It is clear from the over-approximations of these figures, that the good trajectory tracking in the space of $x$ observed in Figures~\ref{fig:theta1}--\ref{fig:omega3}, does not translate well in the space of $y$ under parameter uncertainties. E.g., the bounds for $y_{CoM}\left( t \right)$ in Figure~\ref{fig:yCoM} are up to $\pm 5 \left[ cm \right]$ from its reference trajectory, while $\dot{y}_{CoM}\left( t \right)$ in Figure~\ref{fig:vyCoM} can exhibit deviations of $\pm 2\left[ cm/s \right]$.
	
\begin{figure}[H]
		\centering
		\subfloat[Position trajectories of the three-link robot CoM.\label{fig:CoMposition}]{\includegraphics[width=7.5cm]{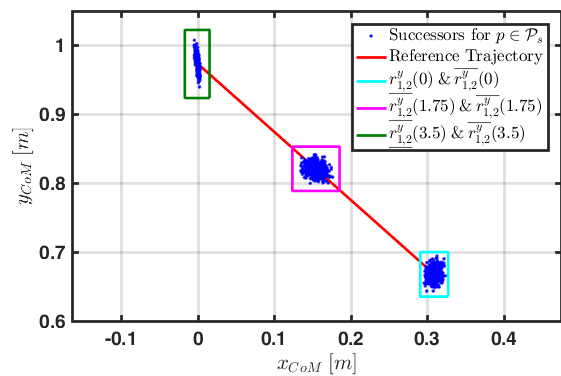}} \quad \quad \quad
		\subfloat[Velocity trajectories of the three-link robot CoM.\label{fig:CoMvelocity}]{\includegraphics[width=7.5cm]{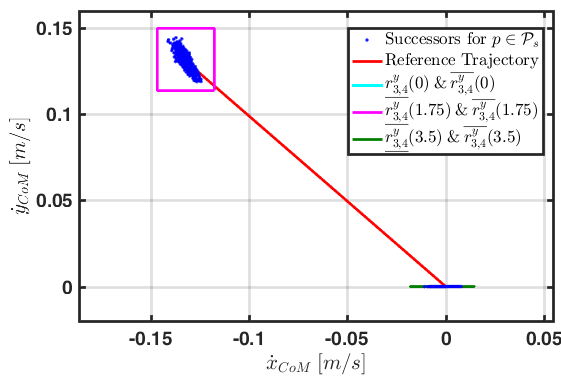}}
		\caption{Over-approximations for the CoM trajectories at three time instants of the STS movement.\label{fig:CoMTrajectories}}
	\end{figure}

	Figure~\ref{fig:CoMposition} shows the projection of the over-approximation interval $[\underline{r^y}(t),\overline{r^y}(t)]$ for the position of the CoM $\left[x_{CoM};y_{CoM}\right]$ at $t_0=0$ (cyan), $t=1.75$ (magenta) and $t_f=3.5$ (green). The clouds of successors $\zeta(\Phi^x(t;0,x_0,p),p)$ from the random parameters $p \in \mathcal{P}_s$ are displayed in blue for each of these three time instants. The nominal trajectory for the whole STS movement is in red. Note that despite having a single initial state $x_0$ for the closed-loop system \eqref{eq:Nonlinear}, the over-approximation $[\underline{r^y}(0),\overline{r^y}(0)]$ at $t_0=0$ is not reduced to a single point, due to the influence of the parameter uncertainty $p\in [\underline p,\overline p]$ on the initial position of the CoM through the mapping $y_0=\zeta(x_0,p)$. The size of the box enclosing the final position of the CoM allows to assess that there is no risk of sit-back or step failures \cite{Eby2006}.
	
	Figure~\ref{fig:CoMvelocity} depicts the projection of the over-approximation interval $[\underline{r^y}(t),\overline{r^y}(t)]$ for the velocity of the CoM $\left[\dot{x}_{CoM};\dot{y}_{CoM}\right]$. The reference trajectory in red goes from $[0;0]$ at $t_0=0$ to $[-0.13;0.13]$ at $t=1.75$ and back to $[0;0]$ at $t_f=3.5$. In this plane, the projection of $[\underline{r^y}(0),\overline{r^y}(0)]$ is reduced to the single state $\{[0;0]\}$ due to the starting conditions at rest $\dot\theta_1(0)=\dot\theta_2(0)=\dot\theta_3(0)=0$. Notice that the projection of $[\underline{r^y}(3.5),\overline{r^y}(3.5)]$ at the final time is almost flat, since $\dot y_{CoM}(3.5)$ goes close to $0$ for every parameter in $[\underline p,\overline p]$, which is beneficial to avoid the feet to lose contact with the ground. 

\begin{figure}[H] 
		\centering
		\subfloat[Torque applied at the hips by the PLLO.\label{fig:tau1}]{\includegraphics[width=7.5cm]{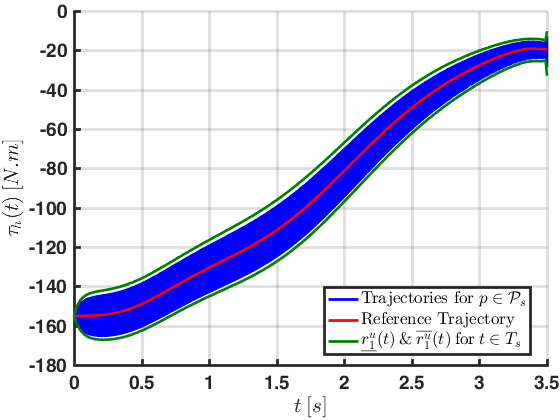}} \quad \quad \quad
		\subfloat[Torque at the shoulders of the user.\label{fig:tau2}]{\includegraphics[width=7.5cm]{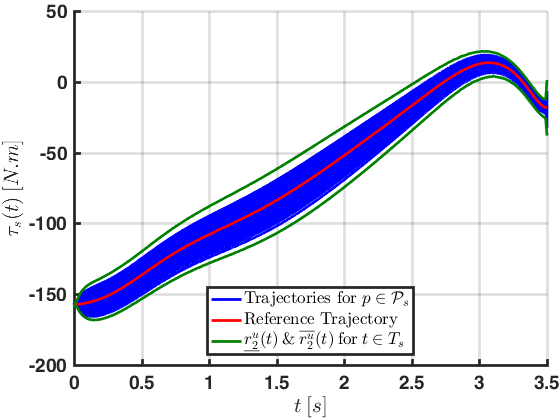}}
		\\
		\subfloat[Horizontal force at the shoulders of the user.\label{fig:Fx}]{\includegraphics[width=7.5cm]{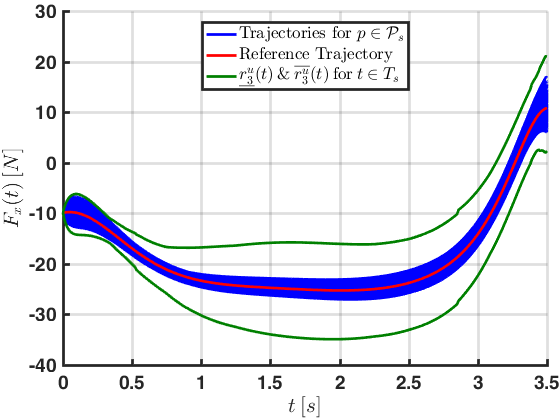}} \quad \quad \quad
		\subfloat[Vertical force at the shoulders of the user.\label{fig:Fy}]{\includegraphics[width=7.5cm]{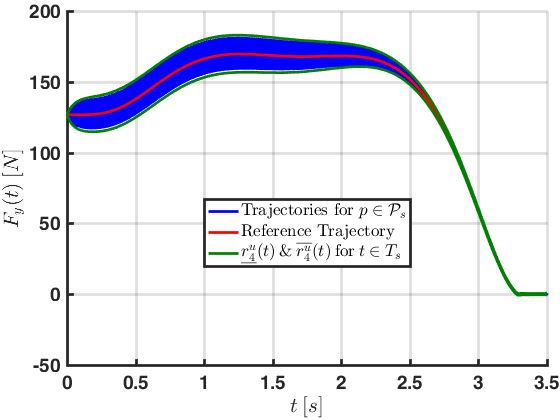}}
		\caption{Input over-approximations $ [\underline{r^u}(t),\overline{r^u}(t)] $ for every $t\in T_{s}$ during the STS movement.\label{fig:Input}}
	\end{figure}

	For the reachability analysis with respect to the control input $u=[\tau_{h};\tau_{s};F_{x};F_{y}]$, we use the state feedback $u(t)=K(t,x(t),p)$ defined by the controller in~\eqref{eq:LQRgain}:{\small \begin{align}
		\label{eq:feedback controller}
		K(t,x(t),p)&:=\hat u(t)-K_{LQR}(t)(x(t)-\hat x(t)).
		\end{align}}The sensitivity $S^{u}$ in \eqref{eq:sensitivity u} can then be reduced to:{\small \begin{align}
		\label{eq:sensitivity u reduced}
		S^{u}\left(t;0,x_0,p\right)&=-K_{LQR}\left(t\right)S^{x}\left(t;0,x_0,p\right).
		\end{align}}
	Applying Proposition~\ref{prop:over approximation} on $\Psi^{u}(t;0,x_0,p)$ with the sensitivity bounds from \eqref{eq:sensitivity u reduced}, allows to compute the over-approximation bounds $\underline{r^u}(t),\overline{r^u}(t)$ shown in green in Figures~\ref{fig:tau1}--\ref{fig:Fy}, alongside the reference trajectory $\hat u(t)$ in red, and the trajectories $\Psi^{u}(t;0,x_0,p)=\hat u(t)-K_{LQR}(t)(\Phi^{x}(t;0,x_0,p)-\hat x(t))$ for the $500$ random $p\in \mathcal{P}_s$ in blue. Since the inputs related to the upper body loads at the shoulders joint are expected to be learnt by the user through training, it is not a good feature of this particular finite time horizon LQR controller that the over-approximations for $\tau_{s} \left( t\right)$, $F_{x} \left( t\right)$, and $F_{y} \left( t\right)$ exhibit deviations of up to $\pm 40 \left[N \cdot m \right]$, $\pm 10 \left[ N \right]$ and $\pm 13 \left[ N \right]$, respectively. Although it could be feasible to apply such loads, the predicted variability with the parameter uncertainty might make it difficult for a user to properly time the actions for a successful ascending phase.
	\pagebreak
	\pagebreak
	Despite applying the reachability analysis with sensitivity bounds estimated from the finite set $\mathcal{P}_{b}$, which are not guaranteed to contain all possible sensitivity values over the parameter interval $[\underline{p},\overline{p}]$, Figures~\ref{fig:theta1}--\ref{fig:Fy} show that all trajectories of \eqref{eq:Nonlinear} with random parameters (in blue) are indeed contained within the computed over-approximations, and are overly conservative only for $F_{x} \left( t \right)$ in Figure~\ref{fig:Fx}.

\begin{figure}[H]
		\begin{centering}
			\includegraphics[width=7.5cm]{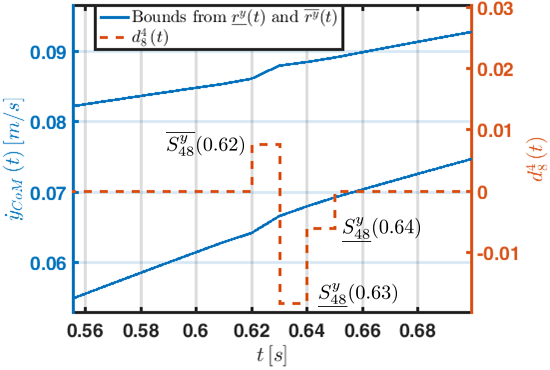}
			\par\end{centering}
		\caption{Effect of $d^i_j(t)$ on over-approximation bounds.\label{fig:CorrectionFactorVyCoMDetail} }
	\end{figure}

	As it can be seen in Figures~\ref{fig:vxCoM} and \ref{fig:vyCoM}, the over-approximations calculated with Proposition~\ref{prop:over approximation} may present non-smooth behaviors.
	This is due to the definition of the compensation term $d^i_j(t)$ in \eqref{eq:param bounds for OA} which may have non-continuous jumps over time between a constant value at $0$ and the sensitivity bound functions {\small $\underline{S_{ij}},\overline{S_{ij}}:[t_0,t_f]\rightarrow\R$}.
	As an illustration, Figure~\ref{fig:CorrectionFactorVyCoMDetail} presents a zoom of Figure~\ref{fig:vyCoM}, where two such non-smooth behaviors are visible on the bounds of the over-approximation (in green) corresponding to the jump from $0$ to $\overline{S^y_{48}}$ at time $t=0.62\left[ s \right]$ and the jump from $\overline{S^y_{48}}$ to $\underline{S^y_{48}}$ at time $t=0.63\left[ s \right]$.

	A workstation of 4 cores at $2.7 [GHz]$ running Matlab Parallel Toolbox completes the sensitivity-based reachability analysis of this section in $5.9[h]$. $1.05[h]$ are spent in solving the sensitivity equation \eqref{eq:sensitivity system} for the set of $500$ $p\in\mathcal{P}_b$. Computing {\small $\underline{S^x},\overline{S^x}:[0,3.5]\rightarrow\R^{6\times 12}$} and {\small $[\underline{r^x}(t),\overline{r^x}(t)]$} take $1.92[h]$, {\small $\underline{S^y},\overline{S^y}:[0,3.5]\rightarrow\R^{4\times 12}$} and {\small $[\underline{r^y}(t),\overline{r^y}(t)]$} take $1.89[h]$, and {\small $\underline{S^u},\overline{S^u}:[0,3.5]\rightarrow\R^{4\times 12}$} and {\small $[\underline{r^u}(t),\overline{r^u}(t)]$} take $1.03[h]$. 
	\section{CONCLUSIONS AND FURTHER WORK}
	
	This paper considered the control problem of the Sit-to-Stand (STS) movement for a Powered Lower Limb Orthosis (PLLO) and its user. A sensitivity-based reachability analysis was applied to evaluate the robustness against parameter uncertainty of a finite time horizon LQR controller.
	Based on the initial computation of lower and upper bounds for the possible sensitivity values over the parameter uncertainty interval, this approach then obtains an over-approximation of the set reachable by the closed-loop system at a given time.
	An extension of this reachability analysis was also introduced to cover auxiliary static systems such as those defined by an output function or the state feedback control.
	
	The over-approximations computed for the PLLO were finally provided in simulations to evaluate the worst-case performances of the system under the control design in~\cite{Narvaez-Aroche2018}. The results highlighted its weaknesses to both track the reference trajectories for the kinematics of the CoM, and guarantee small variations of the inputs at the shoulders joints, by displaying large projections of the reachable sets on these variables. Since the loads on shoulders are expected to be applied by the user with no intervention of the controller, it is desirable to observe small differences between the bounds set by the over-approximations while aiming to minimize the training time needed for the user to perform safe and autonomous STS movements. Future work on this topic will thus exploit the over-approximations of the reachable sets to define a performance metric for choosing a more suitable control strategy.
	
	\section*{ACKNOWLEDGMENTS}
	
	Octavio Narvaez-Aroche would like to thank the Consejo Nacional de Ciencia y Tecnolog\'{i}a (CONACYT), the Fulbright-Garc\'{i}a Robles program and the University of California Institute for Mexico and the United States (UC MEXUS) for the scholarships that have made possible his Ph.D. studies. Andrew Packard acknowledges the generous support from the FANUC Corporation. The authors gratefully acknowledge support from the NSF under grant ECCS-1405413.
	
	\bibliographystyle{asmems4}
	\bibliography{STSBiblio}
	
\end{document}